\documentclass[twocolumn,aps,prd,amsmath,amssymb]{revtex4}
\bibpunct{}{}{,}{s}{}{,}
\usepackage{graphicx,psfrag}   

\def\be{\begin{equation}}
\def\ee{\end{equation}}
\def\ba{\begin{eqnarray}}
\def\ea{\end{eqnarray}}

\newcommand\nn{\nonumber}
\newcommand\q{\quad}
\def\Nl{{\mathchoice
{\setbox0=\hbox{$\displaystyle\rm N$}\hbox{\hbox to0pt
{\kern0.4\wd0\vrule height0.9\ht0\hss}\box0}}
{\setbox0=\hbox{$\textstyle\rm N$}\hbox{\hbox to0pt
{\kern0.4\wd0\vrule height0.9\ht0\hss}\box0}}
{\setbox0=\hbox{$\scriptstyle\rm N$}\hbox{\hbox to0pt
{\kern0.4\wd0\vrule height0.9\ht0\hss}\box0}}
{\setbox0=\hbox{$\scriptscriptstyle\rm N$}\hbox{\hbox to0pt
{\kern0.4\wd0\vrule height0.9\ht0\hss}\box0}}}}
\def\Zl{{\mathchoice
{\setbox0=\hbox{$\displaystyle\rm Z$}\hbox{\hbox to0pt
{\kern0.4\wd0\vrule height0.9\ht0\hss}\box0}}
{\setbox0=\hbox{$\textstyle\rm Z$}\hbox{\hbox to0pt
{\kern0.4\wd0\vrule height0.9\ht0\hss}\box0}}
{\setbox0=\hbox{$\scriptstyle\rm Z$}\hbox{\hbox to0pt
{\kern0.4\wd0\vrule height0.9\ht0\hss}\box0}}
{\setbox0=\hbox{$\scriptscriptstyle\rm Z$}\hbox{\hbox to0pt
{\kern0.4\wd0\vrule height0.9\ht0\hss}\box0}}}}
\def\Ql{{\mathchoice
{\setbox0=\hbox{$\displaystyle\rm Q$}\hbox{\hbox to0pt
{\kern0.4\wd0\vrule height0.9\ht0\hss}\box0}}
{\setbox0=\hbox{$\textstyle\rm Q$}\hbox{\hbox to0pt
{\kern0.4\wd0\vrule height0.9\ht0\hss}\box0}}
{\setbox0=\hbox{$\scriptstyle\rm Q$}\hbox{\hbox to0pt
{\kern0.4\wd0\vrule height0.9\ht0\hss}\box0}}
{\setbox0=\hbox{$\scriptscriptstyle\rm Q$}\hbox{\hbox to0pt
{\kern0.4\wd0\vrule height0.9\ht0\hss}\box0}}}}
\def\Rl{{\mathchoice
{\setbox0=\hbox{$\displaystyle\rm R$}\hbox{\hbox to0pt
{\kern0.4\wd0\vrule height0.9\ht0\hss}\box0}}
{\setbox0=\hbox{$\textstyle\rm R$}\hbox{\hbox to0pt
{\kern0.4\wd0\vrule height0.9\ht0\hss}\box0}}
{\setbox0=\hbox{$\scriptstyle\rm R$}\hbox{\hbox to0pt
{\kern0.4\wd0\vrule height0.9\ht0\hss}\box0}}
{\setbox0=\hbox{$\scriptscriptstyle\rm R$}\hbox{\hbox to0pt
{\kern0.4\wd0\vrule height0.9\ht0\hss}\box0}}}}
\def\Cl{{\mathchoice
{\setbox0=\hbox{$\displaystyle\rm C$}\hbox{\hbox to0pt
{\kern0.4\wd0\vrule height0.9\ht0\hss}\box0}}
{\setbox0=\hbox{$\textstyle\rm C$}\hbox{\hbox to0pt
{\kern0.4\wd0\vrule height0.9\ht0\hss}\box0}}
{\setbox0=\hbox{$\scriptstyle\rm C$}\hbox{\hbox to0pt
{\kern0.4\wd0\vrule height0.9\ht0\hss}\box0}}
{\setbox0=\hbox{$\scriptscriptstyle\rm C$}\hbox{\hbox to0pt
{\kern0.4\wd0\vrule height0.9\ht0\hss}\box0}}}}
\def\Hl{{\mathchoice
{\setbox0=\hbox{$\displaystyle\rm H$}\hbox{\hbox to0pt
{\kern0.4\wd0\vrule height0.9\ht0\hss}\box0}}
{\setbox0=\hbox{$\textstyle\rm H$}\hbox{\hbox to0pt
{\kern0.4\wd0\vrule height0.9\ht0\hss}\box0}}
{\setbox0=\hbox{$\scriptstyle\rm H$}\hbox{\hbox to0pt
{\kern0.4\wd0\vrule height0.9\ht0\hss}\box0}}
{\setbox0=\hbox{$\scriptscriptstyle\rm H$}\hbox{\hbox to0pt
{\kern0.4\wd0\vrule height0.9\ht0\hss}\box0}}}}
\def\Ol{{\mathchoice
{\setbox0=\hbox{$\displaystyle\rm O$}\hbox{\hbox to0pt
{\kern0.4\wd0\vrule height0.9\ht0\hss}\box0}}
{\setbox0=\hbox{$\textstyle\rm O$}\hbox{\hbox to0pt
{\kern0.4\wd0\vrule height0.9\ht0\hss}\box0}}
{\setbox0=\hbox{$\scriptstyle\rm O$}\hbox{\hbox to0pt
{\kern0.4\wd0\vrule height0.9\ht0\hss}\box0}}
{\setbox0=\hbox{$\scriptscriptstyle\rm O$}\hbox{\hbox to0pt
{\kern0.4\wd0\vrule height0.9\ht0\hss}\box0}}}}

\newcommand{\bd}{\mathbf d}

\begin{document}

\title{Diffeomorphism symmetry in quantum gravity models}

\author{Bianca Dittrich}
\email{b.dittrich@uu.nl} \affiliation{Spinoza Institute,
Universiteit Utrecht}
    
\begin{abstract}
    We review and discuss the role of diffeomorphism symmetry in quantum gravity models. Such models often involve a discretization of the space--time manifold as a regularization method. Generically this leads to a breaking of the symmetries to approximate ones, however there are incidences in which the symmetries are exactly preserved. Both kind of symmetries have to be taken into account in covariant and canonical theories in order to ensure the correct continuum limit. We will sketch how to identify exact and approximate symmetries in the action and  how to define a corresponding canonical theory in which such symmetries are reflected as exact and approximate constraints.

\bigskip

\end{abstract}

\maketitle

\section{Introduction}

The dynamics of general relativity is deeply intertwined with space--time diffeomorphisms. This becomes apparent in the canonical formulation where Hamiltonian constraints generate the dynamics and at the same time space--time diffeomorphisms \cite{bergman}. Moreover requiring a canonical theory of spatial geometries and a representation of the Dirac algbra -- the canonical incarnation of diffeomorphisms -- results uniquely \cite{teitel2} in canonical general relativity.

Hence a quantum theory of geometries with a quantum representation of diffeomorpisms should yield a quantum theory of gravity with the correct semi--classical limit. Unfortunately such a representation seems to be out of reach yet. One reason out of many is that many models, such as Regge calculus \cite{regge,rreview} and spin foam models \cite{spinfoams}, involve an explicit discretization of the space--time manifold as a regularization method. Generically this seems to result in a breaking of the continuum diffeomorphism symmetries.  

This can be taken as an advantage as such symmetries lead to divergencies in the path integral and constraints in the canonical formalism and complicate therefore the construction of a quantum theory. The diffeomorphism symmetries can however not be completely ignored as these have to be recovered in the continuum limit. Even in discrete theories there should exist at least approximate symmetries, which should become exact in the limit.

Path integrals for constrained theories such as general relativity can be understood as providing a projector on the space of physical, that is gauge invariant, states of the canonical theory. These are states that are annihilated by the constraints. The heuristic idea is that path integrals implement averagings of the wave functions over the gauge orbits. This can only be made consistent if the symmetries of the canonical and the covariant theory can be brought to match. In particular approximate symmetries in the action should correspond to some notion of approximate constraints.     

We therefore start in section \ref{action} with a review and discussion of symmetries for discretized actions, in particular the Regge action. We will clarify that it is not sufficient to identify infinitesimal variations leaving the action invariant. A further requirement is that the associated transformations have to act non--trivially on the extrema of the action. We will find that in general the symmetries defined in this way are a mixture of exact and approximate ones. In section \ref{triang} we discuss additional symmetries and problems that can arise if one includes a sum over triangulations into the path integral.

In section \ref{canon} we will line out how to obtain canonical formulations that match the symmetries of the discretized covariant theory. To this end we use the method of `consistent discretizations' \cite{consistent} together with a local evolution scheme for the dynamics of triangulated manifolds \cite{commi}. In this way we can make the idea of approximate constraints precise. As a simplified example we will consider 3d Regge calculus with and without cosmological constant in section \ref{mini}. 

Section \ref{dirac} discusses notions of Dirac's algebra for discrete manifolds. In particular we point out that exact versions exist for 3d gravity and for 4d theories describing flat space. Since `chunks of flat space' can also appear in the full 4d case and moreover in the continuum limit curvature should become small compared to the discretization scale such a finding could ensure that an exact 4d Dirac algebra can be recovered in the continuum limit. 

In section \ref{lqg} we review the role of diffeomorphsim symmetries in loop quantum gravity and spin foam models and sketch consequences of our earlier discussions for these theories. We conclude with a discussion and outlook in section \ref{laber}.

\subsection{Regge calculus}

The examples we will consider are taken from Regge calculus. We will therefore give a very short introduction and fix notations. More details can be found in \cite{regge,rreview} and references therein. 

Regge calculus is usually considered on a fixed triangulation $T$ of a space--time manifold. The variables appearing in the Regge action are given by the edge lengths $\{l_e\}_{e \in T}$. This completely specifies the geometry of the triangulation. From the edge length one can compute the 4d dihedral angles $\theta^\sigma_t$ and 3d dihedral angles $\theta^\tau_e$. For a 4--simplex $\sigma$ the dihedral angle $\theta^\sigma_t$ gives the (inner) angle between the two tetrahedra sharing the triangle $t$. A similar definition applies to the 3d angles in a tetrahedron $\tau$ at an edge $e$. 

Consider a triangle $t$ in a 4d triangulation. This triangle is shared by several 4--simplices. A (Levi--Civita) parallel transport of a vector from one 4--simplex to the next around the triangle results in a rotation of this vector by the so--called deficit angle $\epsilon_t=2\pi-\sum_{ \sigma \ni t} \theta^\sigma_t$ in the plane perpendicular to this triangle. The deficit angle measures the curvature concentrated at the triangle. Accordingly the (Euclidean) Regge action with cosmological constant $\lambda$ is given by
\be
S=-\sum_{t \in \text{bulk}} A_t \epsilon_t +\lambda \sum_\sigma V_\sigma
\ee
where $A_t$ denotes the area of a triangle $t$ and $V_\sigma$ the volume of a 4--simplex $\sigma$. We work in units with $c=8\pi G_{Newton}=1$. If there is a non--vanishing boundary the boundary term 
\be
S_{\text{bdry}}=-\sum_{t \in \text{bdry}} A_t (\pi-\sum_{\sigma \ni t}\theta^\sigma_t)  \q 
\ee
has to be added to the action in order to make the boundary value problem (with prescribed edge lengths) well defined. The equations of motion can be obtained by varying the Regge action with respect to the length variables and are given by
\be
-\frac{1}{2}\sum_{t \ni e} \epsilon_t \, l_e \cot \alpha^t_e + \lambda \sum_{\sigma \ni e} \frac{\partial V_\sigma}{\partial l_e} =0
\ee
where $\alpha_e^t$ is the angle in the triangle $t$ opposite the edge $e$. Derivatives of the deficit angles vanish due to the Schl\"afli identity ensuring that $\sum_{t \in \sigma} A_t \delta \theta^\sigma_t=0$ for variations of the dihedral angles in a 4--simplex. This is analogous to the continuum where for the variation of the Einstein--Hilbert action $\int \sqrt{g} g^{ab}R_{ab} d^3x$ the term with the variation of the Ricci tensor leads to a total divergence.

For 3d Regge calculus we have the action
\be\label{3dl}
S=-\sum_{e \in \text{bulk}} l_e \epsilon_e +\lambda \sum_\tau V_\tau +\sum_{e \in \text{bdry}} l_e(\pi-\sum_{\tau \ni e} \theta_e^\tau) 
\ee
yielding the equations of motion
\be
-\epsilon_e+\lambda\sum_{\tau \ni e}\frac{\partial V_\tau}{\partial l_e} =0 \q .
\ee
Here $\epsilon_e=2\pi-\sum_{\tau \ni e} \theta^\tau_e$ is the 3d deficit angle. If the cosmological constant $\lambda=0$ the equations of motion require the deficit angles to vanish, hence in this case solutions are given by flat space triangulations.

\section{Bianchi identities and symmetries of the Regge action} \label{action}

Regge calculus seems to circumvent the problem of choosing coordinates 
and to deal directly with discretized geometries. Hence the title of
Regge's paper \cite{regge} `General relativity without
coordinates'. This however does not guarantee that gauge degrees of freedom will not appear. Also in the continuum the problem is not so much in choosing coordinates, but that different metric fields which seem naively to be expressed in the same coordinates describe the same geometry. 

That gauge degrees of freedom appear in Regge calculus was first shown by Ro\v{c}ek and Williams \cite{rocekwilliams}. In a perturbative expansion of the 4d path integral around flat space non--propagating modes have been identified and it was realized that these correspond exactly to a discretization of the continuous gauge degrees of freedom. These modes have a simple explanation: any triangulation of flat space is a solution of Regge's equation. An arbitrary vertex of such a triangulation can be translated in four directions, keeping the triangulation flat but inducing the length of the adjacent edges to change. These are exactly the gauge modes.



For a general non--flat triangulation the issue is naturally more involved.
Many authors \cite{rocekwilliams,hartle1,miller,moser}
pointed out that there should be at least an approximate notion of
symmetry if the deficit angles (determined by
 the edge length compared to the curvature radius) are small. The
arguments rest on the fact that the (linearized) Bianchi identity
holds approximately, which can be used to show that the Regge
action should be approximately invariant under infinitesimal
vertex displacements. 

Some authors argue even for the existence of exact symmetries of the Regge
action, notably Hamber and Williams\cite{hwgauge}.

Instead of trying to define what the equivalent of a continuous
diffeomorphism might be for a discrete manifold, we will take the
Regge action and hence the discrete system it describes as
fundamental and just ask for symmetries of this action, as is
advocated for instance by Williams \cite{williamsr}. That is we look for
infinitesimal transformation of the length variables that leave the Regge action
invariant. 

Indeed one can convince oneself by a counting argument that for $n$--dimensional triangulations at least
$n$ infinitesimal such transformations per vertex should exist.(A
similar argument applied to 3--dimensional triangulations appeared
in \cite{rocekwilliams}.)
The piecewise
linear manifold conditions for an $n$--dimensional triangulation 
require that every point has a
neighborhood homeomorphic to the open ball in $\Rl^n$.  From this it follows that 
a vertex in such a triangulation is
shared by at least $(n+1)$ edges. Hence there should be at least $n$
variations of the associated length variables that leave the
Regge action invariant.

For the 3--dimensional Regge action without cosmological constant
such variations were found by Freidel
and Louapre \cite{fl} with the help of the {\it exact} Bianchi identities.
 These transformations were actually defined in a first order
formulation, but can be adopted to 3d Regge
calculus. We will repeat the argument here, since it will shed
some light on the 4d case.

To this end consider the simplest case, that is a vertex $v$ that
is shared by $4$ edges $e_i,i=1,\ldots,4$. We will choose some
Cartesian coordinate system, which we can associate for instance
to the tetrahedron $T_1$, that has $e_1,e_2,e_3$ as its edges. The
deficit angles are encoded in parallel transport matrices around
the edges $U_1,\ldots U_4 \in SO(3)$. Here $U_1,U_2, U_3$ start
all in the tetrahedron $T_1$ and are expressed in the Cartesian
coordinates associated to this tetrahedron. With $U_4$ we denote
the parallel transport matrix around $e_4$ starting and ending in
some other tetrahedron but conjugated with the parallel transport
to $T_1$. In this way all parallel transports can be made to start
and end at the same point and are expressed in the same coordinate
system. We will denote the resulting matrices by $U^{ab}_i$.
As one can check easily the composition of the loops
around the edges $e_1$ to $e_4$ can be contracted to a point without 
crossing any edges. Hence the associated parallel transport matrix must be trivial,
that is we have 
\be\label{bianchi1} 
\prod_i U_i :=\prod_i \exp(A_i) = \mathbf{1} \q 
\ee
 where we defined the Lie algebra
elements $A_i:=\log U_i$ (assuming that the deficit angles are so
small that the $U_i$ are in a neighborhood of unity where the
exponential function is invertible). We will need the Bianchi
identity (\ref{bianchi1}) in its logarithmic form, however we have
to face the problem that 3d rotations do not commute. This is the
point where for instance Miller et al.\cite{miller}~apply the `Abelian
approximation' for small deficit angles 
\be \label{bianchi2}
\sum_i A_i \approx 0 \q , 
\ee 
and hence derive an approximate
invariance of the action. Freidel and Louapre note that one can
use the Baker--Campell--Hausdorff formula in order to find the
exact logarithmic expression for (\ref{bianchi1}) in the form
\be
\label{bianchi3} \sum_i \left( A_i + [\Omega_i,A_i]\right)=0 
\ee 
where
$\Omega_i$ are Lie algebra elements depending in a complicated
way on the elements $A_j$ and therefore the deficit angles
$\epsilon_i$. Representing $so(3)$ Lie algebra elements as $B^{ab}=B^c\epsilon^{cab}$ with $\epsilon^{cab}$ the totally antisymmetric Levi-Civita tensor, (\ref{bianchi3}) turns into
\be\label{bianchi32}
\sum_i \left( A_i^c 
+ A_i^d \Omega_i^e \epsilon^{dec} \right)
 \q .
\ee

To find the associated invariance of the action we express the
relevant part of the Regge action with the help of the Cartesian
coordinates introduced before as 
\be \label{bianchi33}
S\propto \sum_i
\epsilon_{abc} A_i^{ab}e_i^{c} 
\ee 
where $e_i^c$ are the edges expressed in the Cartesian coordinates
 and $A_i^{ab}=\epsilon_i \epsilon^{abc}e_i^c/l_i$. (For the edge $e_4$ we use the coordinates in $T_1$.) Variations $\delta e_i^c$ of the
coordinates $e_i^c$ will induce variations in the associated
lengths and we obtain 
\be\label{bianchi34} 
\delta S \propto \sum_i \epsilon^{abc}A_i^{ab}\delta e_i^c  \q . 
\ee 
The terms with $\delta A_i^{ab}=\delta( \epsilon_i) \epsilon^{abc}e_i^c/l_i+\epsilon_i \epsilon^{abc}\delta(e_i^c/l_i)$ vanish. The first summand is orthogonal to $e_i^c$ and for the second summand one can use the Schl\"afli identity.

With the approximate form of the Bianchi identity we obtain that
under the variations \be\label{approx} \delta_{approx} e^c_i :=x^c \ee
describing the displacement of the vertex $v$, the
action is approximately invariant 
\be
\delta_{approx}S= \epsilon^{abc}x^c  \sum_i  A_i^{ab}      \approx 0 \q .
\ee
 But under the variations 
\be\label{exact} 
\delta_{exact} e^c_i
:= x^c -  \Omega_i^{cd}x^d 
\ee the action is exactly invariant: 
\ba
\delta_{exact}S&\propto&\sum_i \epsilon^{abc}A^{ab}_i \left(x^c-\Omega_i^{cd}x^d\right)\nn\\ 
&\propto&  \sum_i 2x^d\left(A_i^d +A^c_i\Omega_i^e \epsilon^{ced}\right) \nn\\
&=&0    
\ea
due to the exact Bianchi identity (\ref{bianchi32}).
Note that the `exact' variations
depend via the Lie algebra elements $\Omega_i$ on the deficit
angles and therefore on the length variables whereas the `approximate'
variations do not. In the case of a flat vertex, where the $\epsilon_i=0$ the two transformations agree.

As the counting argument presented above suggest it should also be
possible to find variations that leave the 3d action with cosmological 
constant term exactly invariant (or any other action for instance 
with higher order terms). For the 4d
Regge action the exact (contracted) Bianchi identities could be
used in a similar way as in 3d to derive invariant variations corresponding
to vertex displacements, for alternative actions again the counting 
argument applies. Indeed invariant variations can be always found.

The problem is however that  the transformations defined by these 
invariant variations might turn out to act
trivially on {\it solutions} of the Regge equations, where 
by definition all variations vanish. 
Consider the example of an
action depending on two variables and invariant under 2d rotations in
these variables. Due to rotational invariance there will be  an
extremum at the center of rotation, hence this solution is a fixed point of the symmetry of the action. There could be
of course other additional sets of solutions with a non--trivial
action of the symmetry.

To test whether candidate symmetries act
non--trivially on solutions one can consider
the Hesse matrix of second derivatives of the Regge action
evaluated on these solutions possesses vanishing eigenvalues or
not. The Hesse matrix also appears in the perturbative treatment
of quantum Regge calculus. For the expansion around flat space
solutions one can indeed find zero eigenmodes corresponding to
vertex displacements \cite{rocekwilliams, hw3d,dfs}. 


To study an (accessible) example with non--flat solutions, we
consider the 3d Regge action with non--vanishing cosmological
constant. 
With the additional volume term the
exact invariance (\ref{exact}) does not need to hold any more,
but as discussed before there
should exist three independent variations per 
vertex that leave the action
(\ref{3dl}) exactly invariant. In the case that these variations
lead to non--trivial transformations also for solutions, we should
find zero eigenvalues for the Hesse matrix evaluated on these
solutions. To simplify the problem as much as possible we consider
the boundary of an equilateral tetrahedron with edge length $a$ and subdivide
this tetrahedron by connecting one inner vertex $v$ to the four
vertices of the tetrahedron by edges with length
$l_i,i=1,\ldots,4$. We have to find the extrema of the action 
as a function of the
$l_i$. There exist one extremum where all the inner edge lengths
are equal $l_i=b$. The length $b$ can be found by solving
 \be\label{eom} 
-(2\pi-3 \arccos \frac{2b^2-a^2}{4b^2-a^2})+\lambda\frac{a^2b}{4 \sqrt{3b^2-a^2}}=0    
\ee
For $\lambda=0$ we obtain $b=\sqrt{3/8}a$. The equation cannot be solved exactly for non--vanishing $\lambda$ but one can obtain a perturbative solution in powers of $\lambda$. 
The $(4\times 4)$ Hesse matrix around a configuration with $l_i=b$ can be
 calculated \cite{dfs} and has a very simple form due to the symmetry of our problem.
All the diagonal entries are equal to a rational function
\be
A=\frac{3}{\sqrt{3b^2-a^2}}\left(  
2\frac{2b^2-a^2}{4b^2-a^2}-\frac{\lambda}{12}\frac{12 b^4-a^4}{3b^2-a^2}
        \right)  \q .
\ee
Also all the non--diagonal elements coincide and are given by
\be
B=\frac{2}{\sqrt{3b^2-a^2}}\left(
-\frac{2b^2}{4b^2-a^2} +\lambda \frac{b^2}{4} \frac{2b^2-a^2}{3b^2-a^2}
\right)  \q .
\ee

 The Hessian matrix
will have three null eigenvalues if and only if $A=B$.

With the equation of motion (\ref{eom}) and the requirement that
$A=B$ we have two conditions for the variable $b$, which can
only be satisfied simultaneously for vanishing cosmological constant
$\lambda=0$. For non--vanishing $\lambda$ the Hesse matrix
evaluated at a solution of the equation of motion will not have
zero eigenvalues, from which we conclude that the candidate
symmetries of the action act trivially on (at least these)
solutions.



In 4d (without cosmological constant) there are zero modes of the Hessian at least for every flat vertex. To the best knowledge of the author there is 
no explicit test for vertices with curvature in the $\lambda=0$ case, but numerical results \cite{galassi} (involving convergence properties of constraints for flat and non--flat solutions) indicate that generically there are no zero modes.

Then different solutions might have symmetry orbits of different
size, according to whether they have flat vertices or not. As
will be explained in the next section, there are many curved
solutions with such flat vertices, moreover the number of such flat vertices might become large for very fine grained triangulations.

Additionally one would expect that with fine graining and hence decreasing deficit angles gauge symmetry will be restored also for the non--flat vertices. In our example the only scale available is the cosmological constant $\lambda$ (with $\lambda=0$ the Regge equations are invariant under global rescalings) hence this corresponds to the $\lambda \rightarrow 0$ limit. The Hessian matrix will exhibit a set of smaller and smaller eigenvalues. The corresponding eigenvectors generate an approximate gauge orbit, along which the action is almost constant.

Note that the inverse of the Hessian appears in saddle point approximations to the path integral. Indeed in the same way diffeomorphism symmetry is gradually restored the path integral should diverge due to the integration over the approximate gauge orbits. Along such an orbit the true extremum (or even several extrema if the solution is not unique) is difficult to locate. Even for classical numerical treatments  solving for the exact extrema of the action becomes so complicated, that instead one (gauge) fixes four length variables per vertex and does not take the corresponding number of equations into account \cite{galassi,commi,kasner}.

\section{Varying the triangulation}\label{triang}

One way to quantize Regge calculus is to fix a triangulation and
to perform the path integral over all allowed length assignments.
One criticism towards this procedure is that the final results may
depend on the choice of triangulation, which is rather an
auxiliary structure. If the triangulation serves as regulator for the 
continuum path integral the induced continuum path integral measure might be 
concentrated around geometries which
arise by assigning roughly the same length to all the edges \cite{david}.

The dependence on one choice of triangulation can be avoided by
summing over all triangulations. That is one considers a
configuration space $C$ labeled by triangulations and length
assignments to the edges of the triangulations (or more generally
some set of  variables on the different k--simplices). Also the
space of triangulations needs to be specified in more detail, for
instance whether one considers fixed topology or not, or whether
the triangulations have to specify the piecewise linear manifold
conditions or not. All these discussions can be summarized in a choice
of measure on the configuration space $C$. Regge calculus on a
fixed triangulation has then only non--vanishing measure on one
triangulation. In contrast the dynamical triangulation measure
\cite{cdt} is concentrated on triangulations with fixed edge
length, that is all the geometric degrees of freedom are put into
the choice of triangulation. Group field theories \cite{gft}
generate sums over topologies, `generalized' triangulations
and variables attached to the simplices of these triangulations.

With the enlargement of the configuration space the question
arises whether some form of diffeomorphism or more general gauge
 symmetries act on this
space involving configurations based on different triangulations.
For instance the authors \cite{roemer,jap} claim that this is
actually the case with Romer et al.\cite{roemer} arguing for a gauge fixing
procedure on this enlarged configuration space.

Indeed for every solution of the Regge equations (with vanishing
cosmological constant) one can introduce trivial refinements of
the underlying triangulation and in this way obtain again
solutions of the Regge equations. These refinements are trivial
since the new edges or triangles do not carry additional
curvature, hence the actual geometry is invariant under these
refinements.

To be more precise, given a solution of the (four--dimensional)
Regge equations choose a 4--simplex $\sigma$ and subdivide this
simplex by placing a vertex $v$ into this simplex. Connect the
vertex $v$ to the vertices of $\sigma$. This introduces five new
edges, ten new triangles and five new 4--simplices. To the new
edges there exist a four--parameter space of length assignments
such that the deficit angles at the new triangles vanish,
corresponding to the possible positions of the new vertex $v$ in
the simplex $\sigma$. Also it is easy to see that the Regge
equations of motion \be\label{reggeb} \sum_{t \ni e} \epsilon_t \cot
\alpha^t_{e}=0\ee are still satisfied: For the new edges the sum
in (\ref{reggeb}) involves only deficit angles of new triangles.
As these are vanishing the equations of motion are satisfied for
the new edges. The equation of motions for the other edges do not
change either, again because the deficit angles at the new
triangles are vanishing.

Hence if one attempts to include all possible triangulations into
the path integral one has to face these redundant configurations.
 One could try to exclude triangulations with trivial subdivisions
from the path integral, for instance by summing only over some class of `simplest' bulk triangulations that interpolate a given boundary triangulation \cite{gftf}. Indeed this works in the (topological) 3d case \cite{fl}. To this end, however, it would be valuable to
know more about the uniqueness properties of the  Regge solutions:
For instance, given a solution with a subdivided 4--simplex, is
this 4--simplex necessarily flat? If this is not the case,
restricting to the `simplest triangulations' could miss out on
physical degrees of freedom. So far some available results
\cite{piran,cp,sub} indicate that solutions to the Regge equations
are not necessarily unique and can vanish or appear under
subdivisions or coarse graining.

A more systematic study of the behavior of Regge solutions under
subdivisions would be valuable, as this could be helpful for obtaining the
continuum limit. If the  conjecture holds, that subdivisions
beyond some scale (determined by either the curvature scale of the
boundary triangulation or the cosmological constant) are trivial, that is only add gauge equivalent solutions,
it might not be necessary to include all triangulations into the
path integral to capture essential physical properties. It might be
even sufficient to restrict to a fixed triangulation, which is fine enough, i.e. solutions should have mostly vanishing deficit angles. Summing over all kinds of
length assignments on this triangulation would include
configurations that are equivalent to geometries defined on coarse grained
triangulations. These terms should
also arise if one chooses to start with another equally fine
triangulation. Hence one would expect physical observables only including scales above the discretization scale to be independent from the concrete choice of this triangulation. 

\section{Canonical theories: Space--time split} \label{canon}

There are several approaches to obtain a canonical or (3+1) version of
Regge calculus, on the one hand in order to develop a numerical
scheme for solving the initial value problem and on the other for canonical 
quantization. The main difficulty in developing
a canonical scheme is firstly the special status of the time variable and
furthermore the question of the existence or non--existence of
gauge symmetries. Both points reappear in some form in loop
quantum gravity and spin foam models, so it might be worthwhile to
discuss these.

There are several options to define a canonical version of
Regge calculus.  One is to start from the discrete
4--dimensional Regge action and to come up with a canonical, that
is initial value formulation that reproduces the 4--dimensional
Regge solutions. This can be summarized as first discretizing and then
 performing a
$(3+1)$--splitting.  Another option is to perform first the
$(3+1)$ splitting and then to discretize. In the latter case the
additional question arises whether to aim for a continuous--time
formulation or an evolution in discrete time steps.

In \cite{wp,friedman} the second viewpoint is adopted, that is to consider
the time evolution of a simplicial 3--manifold in continuous time.
Basically the $(3+1)$ form of the action is evaluated on a three--dimensional
triangulation. The resulting canonical variables are the length of the spatial
edges, conjugated momenta (which in the version \cite{friedman} are in a
non--local relation with respect to the extrinsic curvature leading to a
non--local expression for the Hamiltonian) and lapse and shift functions
associated to the tetrahedra. Accordingly for each tetrahedron there is
one Hamiltonian and three momentum constraints. Whereas in the continuum
these constraints are first class, indicating the gauge freedom to choose
lapse and shift freely, this is not the case for the discretized constraints, which are second class. 

To obtain a consistent evolution in which the constraints are
preserved during time evolution one has to fix\cite{friedman} the values of lapse and shift, hence there is no gauge freedom left. 

That first class constraints turn under discretization into second class ones is a general phenomenon and appears not only in Regge calculus. For instance this happens also for the spatial diffeomorphism constraints\cite{loll} in a lattice version of loop quantum gravity.


As mentioned in the introduction it would be valuable to have a canonical formalism that matches the amount of symmetries inherent in the action. A strategy to ensure this is to construct a canonical formalism that reproduces exactly the solutions of Regge calculus. To this end we consider 4d triangulations and as in a continuum have to choose a slicing of these triangulations into constant time surfaces. Such slices can be defined so that  constant time surfaces are built from the 3d boundaries of the 4--simplices, and such that every vertex of the 4d triangulation is part of at least one constant time surface. Note that there might be many choices of possible slicings for a given triangulation. If the canonical formalism could reproduce all such slicings it would ensure a discrete version of the multi--fingered time picture \cite{kuchar} in continuum general relativity.

 But generically one will encounter the problem that constant time surfaces  in the same triangulation might have a different number of edges and hence would lead to phase spaces of different size. The time evolution map would then have to act between different (sized) phase spaces. It would be of interest to have a formalism that can accommodate such a feature in a canonical formalism, as it might be relevant in particular for cosmological application.

For the time being we restrict ourselves to cases where the 3d triangulation is  the same for every time step. This restriction could lead to very rigid time evolution. Fortunately this is  not the case, as in the so--called puff pastry evolution scheme \cite{commi} an arbitrary 3d triangulation can be evolved in many different ways, so that consecutive 3d triangulations do not change.

That is assume that we have a 4d solution of the Regge equations with a 3d boundary representing the equal time surface. This solution can be evolved further by choosing a vertex $v$ in the boundary to which an edge (the `tent pole') is glued. The other vertex of this edge has then to be connected to the vertices in the 3d boundary adjacent to $v$. In this way a tent is erected and we pushed the hypersurface locally forward in time. Since we can choose (apart from restrictions arising from causality in Lorentzian space-times) at which vertices to evolve the hypersurface this evolution scheme is very reminiscent of a discrete version of multi--fingered time. The question is whether we can find also a continuous or approximate version, that is the freedom to choose lapse and shift at each vertex. Indeed there are numerical indications that this is the case \cite{galassi,kasner}.

Evolving a vertex amounts to solving the Regge equations associated to the new inner edges, that is the pole edge and the edges that are adjacent to the vertex $v$ in the initial hypersurface. These equations have to be solved for the length of the pole edge and the length of the new boundary edges, hence we have as many equations as variables to solve for. The approximate freedom in choosing lapse and shift means that these equations are `approximately' dependent, i.e. values that satisfy all of these but four, will satisfy approximately the remaining four.

Note that these equations for the new inner edges depend on the initial data, that is the edge length in the initial hypersurface and to the past of it. As in the continuum one would expect that these data satisfy constraints (at least in some approximate sense), and only if the initial constraints satisfy these constraints the equations for the new edge length will be approximately dependent. These constraints describe `initial data' on 3d hypersurfaces that can arise as 4d boundaries of Regge solutions.

In the continuum the Hamiltonian and diffeomorphism constraints are given as the contraction of the Einstein tensor with the normal to the hypersurface as $G_{ab}n^a n^b$ and $G_{ab}n^a$ respectively. The Regge equations associated to an edge $e^a$ can be understood as the double projection of the Einstein tensor $G_{ab}e^ae^b$ to this edge. Therefore we will expect that one of the four constraints is given by the Regge equation associated to the tent pole.  

To make this more explicit we have to first find a suitable set of initial data, that is phase space variables that can be associated to a 3d triangulation and then have to be evolved in discrete time steps to reproduce a Regge solution. To this end we can adopt the formalism of consistent discretization \cite{consistent} -- as we will see it amounts essentially to a rewriting of the Regge equations of motion as canonical transformations on a phase space with length and conjugated momenta. The Regge action (with boundary terms) for the 4d tent region serves as a generating function for the canonical transformation implementing the discrete time evolution. This will ensure that the symmetries of the Regge action are reflected in the canonical formalism. 

The consistent discretization scheme has been applied to 3d Regge calculus (with non--vanishing cosmological constant) by Gambini and Pullin \cite{3dcons}. But there a hyper-cubical lattice was chosen making the analysis very complicated. The advantage of the puff--pastry evolution scheme is that we can reduce our considerations to a very small region, in this way we have to deal only with a very small number of equations. A detailed treatment of the 4d dimensional case will appear elsewhere \cite{bahrdittrich}, here we will consider a simplified 3d example corresponding to our earlier discussion in section \ref{action}. As we will see the exact and approximate symmetries discussed there are reflected in the canonical formalism.

\subsection{A mini--triangulation} \label{mini}

\begin{figure}[h]
\includegraphics[width=4cm]{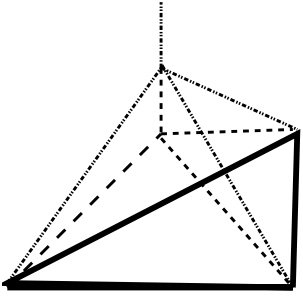}
\caption{\label{fig1} A three--valent vertex is evolved in discrete steps by erecting a pole and connecting the other vertices to the end of the pole. The different types of line indicate the different time steps.}
\end{figure}

The local evolution scheme describes above allows us to resort to a very simple example in order to illustrate the mechanism and the main points. Consider 3d Regge calculus with and without cosmological constant. (The results do not change for the analogous example in 4d). The equal time surfaces in our example will be given by boundaries of tetrahedra, moreover to simplify calculations we require that the three edges of the `base' of the tetrahedron have all length $a$ and the three edges meeting at the `tip' of the tetrahedron have all length $b$. We evolve these surfaces by the evolution described above, that is we join an edge (the pole) with length $t$ to the tip of the tetrahedron and connect the other end of this edge to the vertices of the base, see picture \ref{fig1}.

Hence we have two configuration variables $b$ and $t$, the edge length $a$ does not change in time and therefore is merely a parameter of the system. Performing a step from time $n$ to time $n+1$ three simplices are added that hinge together at the pole. The action (including boundary terms) for this simplicial complex is given by
\ba
S_n(b_n,b_{n+1},t_n)&=& -b_n\, (\pi-2 \theta(b_n,b_{n+1},t_n))\nn\\
&&-b_{n+1}\, (\pi-2\theta(b_{n+1},b_{n},t_n))\nn\\
&&-t_n \,\epsilon(b_{n+1},b_{n},t_n) \nn\\
&&+\lambda   V(b_n,b_{n+1},t_n)
\ea
where $\theta$ are the dihedral angles  at the spatial edges (note that due to the symmetry of our problem we can use the same functions  at two consecutive time steps, but we have to exchange the arguments) and $\epsilon $ is the deficit angle at the pole. The volume of the three tetrahedra is denoted by $V(b_n,b_{n+1},t_n)$. These geometric quantities are given by
\ba
\theta(b,b',t)&=&
{\scriptstyle  
\arccos \left(  \frac{a(b^2-{b'}^2 +t^2)}{\sqrt{4{b'}^2-a^2}\sqrt{2(b^2{b'}^2+b^2t^2+{b'}^2t^2)-b^4-{b'}^4-t^4} }      \right)
} \nn\\
\epsilon(b,b',t) &=&
{\scriptstyle
2\pi-3 \arccos
\left( \frac{2(b^2{b'}^2+b^2t^2+{b'}^2t^2-a^2t^2) -b^4-{b'}^4-t^4}{ 2(b^2{b'}^2+b^2t^2+{b'}^2t^2)-b^4-{b'}^4-t^4   }     \right) 
}   \nn\\
V(b,b',t) &=& 
{\scriptstyle
\frac{a}{4}\sqrt{ 2(b^2{b'}^2+b^2t^2+{b'}^2t^2)-b^4-{b'}^4-t^4-a^2t^2 }
}
  \, .
\ea
We expect the length $t$ of the time like edge to act as the lapse, the shift is vanishing due to our symmetry requirements. 


Following the consistent discretization scheme \cite{consistent} we rewrite the equations of motion for the spatial edge length $b_n$ and the tent pole $t_n$
\ba
0 &=&\frac{\partial}{\partial b_n} \left(S_{n-1}(b_{n-1},b_n,t_{n-1})+S_{n}(b_{n},b_{n+1},t_{n})\right)  \nn \\
0 &=& \frac{\partial}{\partial t_n} S_{n}(b_{n},b_{n+1},t_{n}) 
\ea
 as a canonical transformation given implicitly by
\ba
p_n^b & \equiv &- \frac{\partial S(b_{n},b_{n+1},t_{n}) }{ \partial b_n}  \label{e1} \\
 & =&3(\pi-2 \theta(b_n,b_{n+1},t_n)) - \lambda \frac{\partial V(b_n,b_{n+1},t_n)  }{\partial b_n} \q\q \nn\\
p_n^t & \equiv & -\frac{\partial S(b_{n},b_{n+1},t_{n}) }{\partial t_n} \label{e2}   \\
&=& \epsilon(b_n,b_{n+1},t_n) - \lambda \frac{\partial V(b_n,b_{n+1},t_n)  }{\partial t_n}  \nn \\
p_{n+1}^b &\equiv &\,\, \frac{\partial S(b_{n},b_{n+1},t_{n}) }{\partial b_{n+1}} \label{e3}   \\
& =& -3(\pi-  2 \theta(b_{n+1},b_{n},t_n)) + \lambda \frac{\partial V(b_n,b_{n+1},t_n)  }{\partial b_{n+1}}   \nn \\
p_{n+1}^t & \equiv & \,\, \frac{\partial S(b_{n},b_{n+1},t_{n}) }{\partial t_{n+1}} \label{e4} \\&=& 0   \nn \q .
\ea
Derivatives of the angles vanish due to the Schl\"afli identity.

Equation (\ref{e4}) demanding vanishing of the $t$--momentum, is a constraint. It corresponds to the continuum result that the momentum conjugated to the lapse is vanishing. As in the continuum requiring that this constraint is preserved in time results in a secondary constraint, namely that the right hand side of (\ref{e2}) should be equal to zero. In the continuum this secondary constraint is the Hamiltonian constraint.

To obtain a constraint on the canonical data with the same evolution label $n$, we have to solve equation (\ref{e1}) for $b_{n+1}$ as a function of $b_n,p^b_n$ and $t_n$ and to use this expression in (\ref{e2}). 
\be\label{con1}
C_\lambda(b,p^b,t)=C_0(b,p^b)+\lambda F_\lambda(b,p^b,t) =0\q .
\ee 
In the case $\lambda=0$ this constraint does not depend on $t$
\be\label{con2}
C_0(b,p^b)=2\pi-3 \arccos \left(   \frac{ \cos(\tfrac{p^b}{3})(4b^2-a^2)-a^2}{4b^2}      \right) \q .
\ee 
That the constraint depends only on $b$ and $p^b$  expresses the dynamical fact that the equal time surfaces always bound a flat tetrahedron. This tetrahedron might be subdivided into smaller ones, but since the deficit angles have to vanish the details of the subdivision do not matter. The momenta conjugated to the variable $b$ are given by the dihedral angles of this tetrahedron, which can be computed knowing only the edge length $b$ (and the edge length $a$).

The constraint (\ref{con1}) is preserved under time evolution, that is if $C(b_{n},p_{n}^b)$ vanishes this holds also for $C(b_{n+1}(b_n,p^b_n,t_n),p_{n+1}^b(b_n,p^b_n,t_n  )   )$ with $t_n$ arbitrary. We obtain a consistent evolution, where $p^b_n$ is constrained as a function of $b_n$ but the lapse $t_n$ can be chosen arbitrarily at every time step.


As in the continuum where the lapse function can be dropped from phase space we can omit the variable $t$ and are left with the edge length $b$ and conjugated momentum $p^b$. There is one constraint between the variables generating a one--dimensional gauge orbit. Hence the number of physical degrees of freedom is zero.  The same result holds for arbitrary complicated triangulations (of spheres) in 3d gravity without cosmological constant \cite{dittrichfreidel} and can also be extended to four dimensions for a special class of hypersurface triangulations leading to a dynamics of flat 4--dimensional space \cite{dittrichryan}, see the discussion in the next section. 

The time steps $t_n$ can be chosen arbitrarily, that is also arbitrarily small. In this way we can recover a continuous time formalism in which the constraint (\ref{con2}) does generate time evolution with arbitrary choice of lapse. This reflects the continuous group of symmetries acting on the space of solutions of the covariant theory.

For non--vanishing cosmological constant the constraint (\ref{con1}) does depend on the variable $t_n$. An analytical solution is not possible for $\lambda\neq 0$, but one can make a perturbative ansatz $b_{n+1}= {}^{(0)}b_{n+1}+\lambda{}^{(1)}b_{n+1}+\ldots $ for the solution of equation (\ref{e1}) and use it to obtain the constraint (\ref{con1}). The constraint depends on $t_n$ starting with the quadratic order in $\lambda$. 
Given some initial data $b_0$ and $p_0^b$ one can solve the constraint for $t_0$ and evolve to the next time step where again one has to solve the constraint for $t_1$ and so on. The time intervals $t_n$ are therefore determined by the chosen initial data $b_0$ and $p_0^b$.
This allows to define a reduced system by solving the constraints at all time steps for the time intervals $t_n$ as functions of $b_n$ and $p_n^b$ and to insert these solutions in the remaining time evolution equations. This has the advantage of getting rid of the constraints as well as some of the technical and conceptual difficulties connected to these. This `consistent discretization' approach has been proposed and developed by Gambini,Pullin and collaborators \cite{consistent}. 

In particular the application of this scheme to 3d Regge calculus (with cosmological constant) has been outlined in \cite{3dcons}, choosing a rigid slicing based on a hyper-cubical lattice. Note however that this strategy does not work in instances where the discrete action exhibits symmetries, as in the case $\lambda=0$ (even partly in four dimensions). For 3d gravity the question arises how to obtain the continuum limit, which has a finite number or even zero physical degrees of freedom, starting with a discrete theory with many physical degrees of freedom. Moreover this limit would involve an increasing number of edges and in the scenario without constraints therefore degrees of freedom. 

On the other hand from the discussion of approximate symmetries one would expect that these symmetries should become exact in the continuum limit, with the constraints reappearing in some way. This indicates that choosing the (spatial) edge lengths and conjugated momenta as free variables by solving for the time like edge lengths may lead to a bad parametrization of the dynamically allowed configurations. The constraints (\ref{con1}) could either have no solutions $t_n$ at all or lead to rather large edge lengths $t_n$, spoiling the continuum limit. There is only a small range of variables $(b,p^b)$  that will cause small time intervals.

For this reason we advocate to keep the constraints explicitly and use them as a controlling device for the continuum limit. That is at the initial time step we keep $t_0=\epsilon$ as a free parameter and find initial values $b_0$ and $p^b_0$ satisfying the constraint. These data have to be evolved using the equations (\ref{e1}-\ref{e4}). The difference to an exact reparametrization invariant theory (as the $\lambda=0$ case) is that now $b^n,p_n^b$ {\it and} $t_n$ are fully determined by the evolution equations for $n>0$. In the case that $t_n \sim \epsilon$, that is if the time intervals remain small, a (time) continuum limit $\epsilon \rightarrow 0$ can be obtained. Numerical evolution schemes \cite{galassi,kasner} for Regge equations follow an opposite strategy in choosing the equivalents of lapse and shift freely and ignoring in every time step four equations of motion which are therefore treated as constraints. One hopes that the violation of these constraints does not grow during evolution and tends to zero in the continuum limit. Indeed numerical evidence for this scenario has been found. In this example the constraints are implemented exactly and the question is how the time steps grow. The same numerical evidence can be taken as an indication that if the initial data are chosen carefully, these remain bounded through evolution, although this deserves further study.

To keep the constraints explicitly is closer to the continuum time canonical formalisms with dynamically determined lapse and shift mentioned before and to the viewpoint developed in the work \cite{huwi} which argues that discretizations will in general induce complicated gauge fixings. But the examples considered in this paper differ from the approach followed here. The work \cite{huwi} starts directly with a Hamiltonian formalism in which there are explicit constraints apriori not involving the time intervals $t_n$. Evolution of these constraints under a finite canonical transformation leads however to the appearance of $t_n$ in the time evolved constraints. Conservation of the constraints in time require the $t_n$ to be (gauge) fixed. Here we start from a discrete action and end up with constraints $C(t_n,b_n,p^b_n)$ depending on the time intervals directly. 

We want to stress that this is the main difference to a continuous time canonical formalism, in which one assumes exact constraints, independent of lapse and shift. The discrete time scenario makes the notion of `approximate constraints' precise: we can see these as a family of exact constraints labeled by Lagrange multipliers. Choosing these Lagrange multipliers in a small interval (around zero)  we obtain as allowed configurations for the remaining phase space variables a `thickened constraint hypersurface'. If in the continuum limit the approximate symmetries are restored also the constraints should become exact: One can keep either the range for the Lagrange multipliers fixed, then the `thickness' of the constraint hypersurface should decrease. Or we keep the `thickness' constant, then the range of Lagrange multipliers should increase until the full freedom of their choice is restored.

In the quantum theory it is therefore consistent to impose the constraints only in an approximate sense. For instance in a path integral the integration over the Lagrange multipliers should result in an exact projector on the constraints only if there is a solution (saddle point in a semi-classical approximation) to every choice of Lagrange multiplier. A path integral with approximate symmetries can only impose constraints approximately on the canonical states.

\section{Discrete hypersurface deformation algebras} \label{dirac}

Dirac's hypersurface deformation algebra
\ba\label{alg}
\{H[M],H[N]\}&=&D[g^{ab}(M\partial_b N-N\partial_b M)] \nn\\
\{D[M^a],H[N]\}&=&H[M^b\partial_b N ] \nn\\
\{D[M^a],D[N^b]\}&=& D[M^b\partial_b N^a-N^b\partial_b M^a]
\ea 
can be understood as the canonical incarnation \cite{bergman} of the space--time diffeomorphism group. Here $H[N]=\int_\Sigma N(\sigma) H(\sigma)\bd \sigma$ is the Hamiltonian constraint integrated with a lapse function $N(\sigma)$ over a spatial hypersurface $\Sigma$ whereas $D[N^a]=\int_\Sigma N^a H_a(\sigma)\bd \sigma$ are the spatial diffeomorphism or vector constraints integrated with the shift vector field $N^a(\sigma)$. The algebra is very complicated as we have in the first relation of (\ref{alg}) a structure function appearing that involves the inverse $g^{ab}$ of the spatial metric. 

The hypersurface deformation algebra is describing a universal \cite{teitel} relationship between infinitesimal deformations of a hypersurface in the embedding space. The Hamiltonian constraints generate deformations of the hypersurface normal to itself whereas the vector constraints generate tangential deformations, that is spatial diffeomorphisms. The universality, i.e. independence from the field content of the theory, follows from the fact that the algebra can be derived by purely geometric considerations \cite{teitel}. We want to stress that therefore such a notion of hypersurface deformation algebra should also exist for discretized geometries. Also the structure constants and structure functions should be derivable by geometrical considerations. Another question is whether it has universal properties and whether there are classical and quantum representations of this algebra realized by phase space generators and operators respectively.

It has been shown \cite{teitel2}, that under certain assumptions including the nature and number of phase space variables, canonical general relativity provides the unique phase space representation of the algebra (\ref{alg}). A quantum representation of this algebra could therefore ensure the correct classical limit of the theory. Because of the intricacies and unusual form of the algebra this however has not been achieved so far for any theories, even free parametrized field theories\cite{madha} in space--time dimension larger than two.  

Given the significance of Dirac's algebra for the continuous theory it would be valuable to have a notion of a hypersurface deformation algebra for discrete geometries \cite{bander}. Also here two different notions are possible, one with continuous and one with discrete time or deformation parameters. The next question is to construct (at least classical) representations, that is canonical theories exhibiting these symmetry algebras. Of course it would be interesting to know in which sense the universality features and the uniqueness proof can be carried over to the discrete cases.

For continuous deformation parameters the easiest case to consider are hypersurface deformation algebras describing embeddings of piecewise linear hypersurfaces into flat space.  As in the continuum the algebras itself can be defined in any dimension by purely geometric considerations, that is by deriving the commutation relations of infinitesimal deformations of such hypersurfaces. One question would be how to define such deformations. The most elementary notion (and the one matching the symmetries encountered so far in covariant Regge calculus) is to use vertex displacements, i.e. to change the position of one vertex in the embedding space and to accordingly adjust the length of the adjacent edges. Displacements of higher dimensional simplices can be combined out of vertex displacements.

In the continuum Dirac's algebra is defined by using two types of geometric deformations, the (normalized) deformations normal to the hypersurface and tangential ones. Such a splitting is ambiguous at a vertex in a piecewise linear geometry, so some choice for a normal to the vertex has to be done. The final hypersurface deformation algebra will depend on this choice, details will appear in \cite{dittrichfreidel}.

Note that even a tangential vertex deformation will change the geometry of the spatial hypersurface (with the exception of intrinsically flat vertices), whereas in the continuum the spatial diffeomorphism constraints do not change the intrinsic geometry. Nevertheless for instance for three--dimensional gravity all three vertex deformations are realized as constraints and are needed to obtain the correct counting of physical degrees of freedom for 3d gravity.  The naive viewpoint that the spatial triangulation represents the geometry of the hypersurface, i.e. that the spatial diffeomorphism have been already factored out completely, leads to an incorrect result, see also the discussion in \cite{zapcor}.

This leads us to the next question, namely if there exist any representations of these algebras. In three dimension such representations are provided by several canonical formulations of discretized three--dimensional gravity \cite{waelbroeck3d,louaprefreidel,dittrichfreidel}, which can be understood as the dynamics of hypersurfaces embedded into flat 3d space. One formulation for (Euclidean) 3d gravity is in terms of $SU(2)$ connection variables, see for instance \cite{louaprefreidel}. The phase space variables are given by $SU(2)$ holonomies and triad variables associated to the edges of the triangulations. The dynamics is given by two sets of first class constraints, the Gau\ss~ constraints and the so called flatness constraints. The Gau\ss~ constraints can be used to reduce the phase space to $SU(2)$ invariant variables giving the length of the edges and conjugated momenta providing a canonical formulation of 3d Regge calculus\cite{waelbroeck3d,dittrichfreidel}. (The example in section \ref{mini} with vanishing cosmological constant is a symmetry reduction of these theories.) The flatness constraints generate vertex displacements. A priori the flatness constraints are Abelian and do not have a geometric invariant meaning as normal or tangential generators. Such a split can however be introduced by using appropriate linear combinations (with phase space dependent coefficients) of the flatness constraints. The new algebra is still first class and reflects the geometric commutation relations of the infinitesimal deformations. 

In four dimension a representation for such an algebra can be defined for a restricted class of spatial three--dimensional triangulations \cite{dittrichryan}. This subclass can be understood as a sub-sector of Regge calculus, whose triangula
tions only allow flat solutions (see also the papers \cite{waelbroeckzap}  for defining a flat geometric sub-sector in a canonical formulation of BF--theory). The reason for the restricted class of triangulations is the following: A 2d triangulation with prescribed edge lengths can be generically (ignoring triangle inequalities, self intersections and similar issues) embedded into 3d flat space. This reflects the dynamical content of 3d gravity: Given a 2d metric the constraints can be solved (locally in phase space) for the momenta which define an embedding of the corresponding surface in flat 3d space. 

A generic 3d triangulation with prescribed length cannot be embedded into flat 4d space, as this would require constraints on the length variables \cite{dittrichryan}. However there is a special class of triangulations, such as the boundary of a 4--simplex, which can be embedded into 4d flat space. More complicated triangulations with this property can be generated by applying $(1-4)$ Pachner moves to this triangulation, i.e. by placing a vertex into a tetrahedron, subdividing it thus into four new tetrahedra. A more complete discussion can be found in the paper \cite{dittrichryan} here we will just give the canonical theory describing the `dynamics' of a 4--simplex boundary embedded in flat space.

The canonical variables are given by the ten edge length $l_e$ denoting vertices and conjugated momenta $p_e$. A canonical transformation can be performed (in local patches of phase space) to area variables $A_t(l)$ and conjugated momenta $p_t=\sum_e \frac{\partial l_e(A)}{\partial A_t} p_e$. The dynamics is then defined by the ten constraints
\be\label{triangle}
C_t:=p_t - \theta_t(l) \q 
\ee
where $\theta_t$ is defined to be the dihedral angle at the triangle $t$ in a flat 4--simplex. Due to the Schlaefli identity the constraints are first class and hence generate gauge transformations. These gauge transformations correspond to the vertex deformations of the 4--simplex. Since there are 5 vertices in a 4--simplex we have 20 possible deformations, however there are 10 combinations corresponding to rigid translations and rotations of the 4--simplex which do not change the geometry of the 4--simplex.  

Taking the following linear combinations of constraints
\be
C_e:=\sum_t \frac{\partial A_t(l)}{\partial l_e} C_t=p_e-\sum_t \frac{\partial A_t(l)}{\partial l_e} \theta_t (l)
\ee
we obtain generators that change only one length of one edge at a time (and therefore commute). The constraint $C_e$ induce therefore a displacement of either of the vertices in the direction normal to the other edges adjacent to that vertex. (The ambiguity in associating a vertex stems from the fact that a transformation at both vertices amounts to a rigid transformation of the 4--simplex). Taking linear combinations of these constraints we can define generators leading to arbitrary vertex displacements and compute a deformation algebra. 

Note that in equation (\ref{triangle}) the triangle areas seem to appear as an easier set of variables to deal with. The conjugated momenta are directly related to the dihedral angles, encoding the extrinsic curvature of the triangulated 3d surface. Indeed also in loop quantum gravity areas and 3d dihedral angles between triangles arise as the simplest kind of geometrical variables. These variables provide however an over-complete set and need therefore to be constrained in order to yield consistent length assignments to the edges of the triangulations. Fortunately such constraints are available for 4d triangulations \cite{areaangle} and can be used to define an area--angle Regge action. The same constraints arise for 3d triangulations \cite{dittrichryan} and can therefore be used in a canonical analysis of the area--angle Regge action \cite{areaangle}. The advantage of such a formalism is its closeness to spin foams and loop quantum gravity. In fact the variables used there are related by a Gau\ss~ constraint reduction \cite{dittrichryan} to the area and angle variables.

The incidences where one can define representations of an (infinitesimal) hypersurface deformation algebra correspond exactly to the known appearances of non--trivial symmetries in the corresponding covariant versions, that is Regge calculus. 
If there are no corresponding symmetries in the action, it is questionable that a first class constraint algebra reflecting infinitesimal hypersurface deformations of discretized hypersurfaces can be defined exactly, as this would generate solutions not present in the covariant theory. It might exist in an approximate sense, that is reflecting the algebra (\ref{alg}) modulo anomalous terms vanishing in the continuum limit. This has been shown for the spatial diffeomorphism constraints in a lattice approach to loop quantum gravity \cite{loll}.

In a quantum theory a deformation algebra might rather involve finite transformations. Since the spectra of spatial geometric quantities in loop quantum gravity are discrete \cite{lqgdiscrete} there might appear a similar discreteness in time direction. In the examples discussed so far the choice at which vertices to apply a tent move in the puff--pastry evolution scheme \cite{commi} can be understood as choosing lapse to be either equal to zero or equal to some dynamically determined value. In a quantization such a move could give rise to a finite (unitary) evolution instead of a Hamiltonian viewed as infinitesimal generator. Similarly the spatial diffeomorphism constraints in loop quantum gravity do not exist as infinitesimal generators but only as finite unitary operators.

This understanding of the hypersurface deformation algebra corresponds to different choices of triangulations in the covariant theory. The puff pastry evolution scheme will generate a certain subclass of triangulations. A complete equivalence can be obtained if one considers all the allowed Pachner moves in the $(n-1)$--dimensional hypersurface. These can be related to gluing or removing $n$--simplices from the hypersurface thus building up an $n$--dimensional triangulation \cite{marzouli,dittrichfreidel}. The discrete time notion of hypersurface deformation algebras therefore depends on a better understanding of any symmetries associated with the choice of triangulations.

\section{Loop quantum gravity and spin foams} \label{lqg}

In this section we will consider implications of the results collected in the previous section for quantum gravity models, in particular spin foams and canonical loop quantum gravity \cite{lqg,spinfoams}.
 
Similar to quantum Regge calculus spin foam models  are based on a path integral over discretized geometries. What differs from (traditional) quantum Regge calculus is the choice of variables and the discrete integration measure over these variables . This leads to a sum over spin labels ($SU(2)$group representations) and group intertwiners, associated to subsimplices of the triangulation. The discrete integration measure is motivated among other things by canonical loop quantum gravity, where the spectrum of kinematical geometric operators is discrete and encoded in such spin labels and intertwiners. To obtain the amplitude associated to a given set of labels one considers the (discretized) Plebanski action \cite{plebanski} that formulates general relativity as a second class constrained $BF$ theory. The constraints are imposed by different methods \cite{newmodels} onto the quantum amplitudes. Furthermore the amplitudes depend on a choice of discretized measure which is often left open to some extend in the definition of spin foam models.

 So far the strategy to obtain a semi--classical description is to consider a fixed triangulation and to consider the large spin limit. This is a limit where the discretization scale given by the spectra of geometric operators, i.e. the Planck length becomes small compared to edge length. (There is a second discretization scale defined by the underlying piecewise linear triangulation, in other words the typical edge length.) In this limit one hopes to find an effective description given by the Regge action and indeed there are indications that certain models satisfy this criterion \cite{conrady}. In so far one can understand spin foam models as a quantization of Regge calculus. For other models constructed in a more direct way from first order versions of Regge's action see the work \cite{oriti1}. The question arises whether the exact or approximate symmetries of the discretized action are reflected in spin foam models, that is in the amplitudes defined by the action and the choice of measure.

There is another viewpoint \cite{pro,karim} of the path integral which makes it absolutely necessary to study the symmetries of the path integral: A path integral for constrained theories such as gravity can be understood as a projector onto physical states, i.e. wave functions that satisfy the Hamiltonian and diffeomorphism constraints. (This is not a proper projector as it will map normalizable wave functions to unnormalizable ones.)

The heuristic idea behind is that the path integral implements an averaging over the gauge orbits generated by the constraints, resulting in wave functions that are constant along these orbits. This can only work, if the path integral respects the symmetries, in particular if the measure is gauge invariant. So far this idea is rather heuristic for 4d gravity and has been made only rigorous for 3d gravity \cite{karim}, where the path integral has all the necessary symmetries \cite{fl}.  

As was mentioned before a discretized path integral might have a complicated symmetry structure: diffeomorphism or translation symmetries associated to trivial vertices can be implemented exactly, for non--trivial vertices there might be only approximate symmetries. Here it would be valuable to know whether this can lead at least to an approximate `projector', that is a Gaussian compared to a delta function. This would be consistent to the discussion in section \ref{mini}, where we pointed out that in a discrete theory the allowed initial data are described by a `thickened' constraint hypersurface. In the continuum limit, in which the symmetries become exact, the projector should also become exact. 
Approximate symmetries are reflected by very small eigenvalues in the Hessian of the action, which should converge to zero in the continuum limit. The integral over the directions corresponding to these eigenvalues is an integral over the Lagrange multipliers parameterizing the (approximate) gauge orbit. If the action and the path integral measure is approximately constant along these gauge orbits it should result in a projector onto approximately gauge invariant states.
This behavior should be studied in toy models. Other ideas how to construct a projector on physical states will be mentioned below.

So far we discussed the symmetries of the action, but to obtain a `projector onto physical states' also the path integral measure has to respect the symmetries. In the context of spin foam models this has been discussed extensively by Bojowald and Perez \cite{bojoperez}. They derive two kinds of requirements. The first requirement can be understood to arise from the continuum diffeomorphism symmetry leading to vertex translations in the discretized action. Together with the second class constraints appearing in the Plebanski action the constraints associated to the vertex translations (either second or first class) form a complicated constraint algebra, which has been so far only computed in the continuum case \cite{sergej}. This constraint algebra is however needed to compute the Fadeev--Popov determinant to obtain the correct path integral measure. 

The second requirement is related to symmetries which involve a change of triangulation and labels (so called moves). In spin foam models labels are allowed to have zero spin leading to vanishing edge length -- therefore there are more moves that leave the encoded state unchanged. The amplitudes associated to two configurations related by such moves should be the same. This leads to a number of consistency conditions for the amplitudes, in particular restricting the choice of face and edge amplitudes, that are often left open to choice.

Furthermore Bojowald and Perez argue that a consistent choice of amplitudes has to lead to divergencies. This has to be expected as the path integral includes a sum over non--compact gauge orbit. For the 3d spin foam model this has been discussed by Freidel and Louapre \cite{fl} and the symmetries gauge fixed. For the 4d case, if the spin foam model reproduces the symmetries of the Regge action, one would expect at least divergencies associated to the exact translation symmetry of trivial vertices, whereas approximate symmetries associated to non--trivial vertices could still lead to finite results (however diverging in the continuum limit). Here trivial vertices are those for which saddle points in the Regge action exist such that the deficit angles at adjacent triangles vanish. This depends on the boundary conditions for generic vertices but will always be the case for five--valent vertices.

Spin foam models are hoped for to provide a projector on diffeomorphism invariant states in canonical loop quantum gravity. On the other hand such states are defined to be annihilated by the diffeomorphism and Hamiltonian constraints, and we will discuss certain aspects of these constraints in the remainder of this section.

Loop Quantum gravity provides a rigorous construction of a kinematical Hilbert space which carries a representation of (kinematical \cite{dt}) geometric operators. As was already mentioned these turn out to have a discrete \cite{lqgdiscrete} spectrum, which is one reason for the choice of a discrete path integral measure for the spin foam models. This makes it very tempting to associate a discretized piecewise linear geometry to the eigenstates, the so--called spin networks, of these operators, a picture close to the construction of spin foam models as discretized path integrals. 

But loop quantum gravity is a quantization of the continuum theory and does {\it not} proceed by considering discretized classical geometries and then quantizing. Hence the Hilbert space spanned by spin networks is the state space associated to continuum geometries. As such one would expect the implementation of the continuum constraints and continuum Dirac algebra. Indeed the celebrated FLOST uniqueness theorem \cite{flost}, showing that loop quantum gravity provides a unique kinematical set--up satisfying certain assumptions, requires a representation of the continuous (more specifically semi--analytic) spatial diffeomorphisms.  

Another route is to first discretize the geometry, obtain a canonical description and then quantize using loop quantum gravity techniques, see for instance \cite{mak,immirzi,waelbroeckzap,zapatapiecewise,uniform,rovcosmo,dittrichryan} where this approach is followed or advocated for. Here one would expect the implementation of the (possibly second class) constraints for a discretized geometry. The relations on the kinematical and dynamical level between these two approaches are not quite clear \cite{dittrichryan} and should be explored in more detail.  

As spin foam models are derived from a discretized path integral one would expect that boundary states of this path integral are states obtained from the second route of canonical quantization. Also at least before the continuum limit spin foams should project onto the physical states defined by the constraints in the explicitly discretized theory. Group field theories generate sums over all possible triangulations (or more general complexes), in a certain sense these theories should include information about continuum geometries and dynamics. Hence a canonical formulation of these could be connected to a continuum canonical theory related to loop quantum gravity. The immediate difficulties in obtaining such canonical formulations as well as its potential significance are sketched in the report \cite{oriti}. The full set of symmetries  in the covariant version has still to be identified.

For the continuum theory of loop quantum gravity a uv--finite quantization of the Hamiltonian constraints has been obtained by Thiemann \cite{qsd}. The status of the quantum Dirac algebra in the (continuum) loop quantum gravity is however shrouded by several severe technical difficulties \cite{qsdalg,alglewand,master}. That is the algebra of the Hamiltonian constraints is anomaly free, even Abelian. Abelianess does not reproduce the constraint algebra (\ref{alg}) but is consistent as the Hamiltonian constraints require for their continuum limit the topology provided by the Hilbert space of spatially diffeomorphism invariant states. On this space the diffeomorphism constraints (on the right hand side of the classical Poisson bracket between two Hamiltonians) vanish by definition. A further difficulty in obtaining an infinitesimal algebra is that the spatial diffeomorphism constraints are not quantized as infinitesimal generators but are only implemented via  a family of non--weakly continuous unitary operators. For extended discussions of these aspects we refer to \cite{qsdalg,alglewand}.  

For the second route of first discretizing and then quantizing there is not a full classical definition of the constraint algebra (apart from the sub-sector yielding flat dynamics) yet, but this might be obtained with the techniques described in section \ref{mini}. As discussed there the results might be a complicated mix of second and first class constraints. Additionally one would expect that with decreasing discretization scale the approximate symmetries become exact and hence the second class constraints converted into first class. Before quantization one should test whether this is indeed the case, as otherwise `classical anomalies' are bound to appear.

If one follows the approach to quantize the discrete theory and then to perform the continuum limit the question arises of how to deal with the constraints. Usually second class constraints have to be taken care off at the classical level as they cannot be implemented exactly on quantum states: The uncertainty relations require the fluctuations of the second class constraints to be of the order of the expectation value of the commutator. Again one might hope that heuristically the expectation values of this commutator should become small with the discretization scale (encoded in the state used to obtain the expectation value).  

To solve the second class constraints at the classical level and to introduce Dirac brackets would be overly complicated and might not allow for a continuum limit. Therefore methods are asked for that can deal with mixed constraint systems on the quantum level. There are several such proposals \cite{klauder,uniform,master} based on different motivations. But they have in common to consider just one (`master') constraint, typically a sum or integral of the squares of the original constraints. This constraint is used to construct a projector on the physical Hilbert space, either via path integral methods \cite{klauder} or via spectral analysis. In the latter case, since the master constraint operator has to be positive definite, a physical Hilbert space can be defined as the (generalized) eigenspace corresponding to the minimum of the spectrum. (Alternatively one can consider the generalized eigenspaces from a small interval around zero and in this way obtain a proper projector. This would correspond to the `thickened constraint hypersurface' picture from section \ref{mini}.) As there are second class constraints this minimum will not be zero and may or may not \cite{master} converge to zero in the continuum limit. The `uniform discretization' method \cite{uniform} has been applied to a spherically reduced model with spatial diffeomorphism constraints and it is found that the minimum of the spectrum converges to zero \cite{emergent}.  

Following these approaches a physical Hilbert space can be constructed even with second class constraints. The question arises how to detect quantum anomalies, i.e. to test whether the (hypothetical) quantized second class constraints become first class in the continuum limit. The minimum of the spectrum going to zero in the continuum limit might be one indication but there are examples \cite{master} based on non--compact Lie groups in which the minimum is non--zero despite a non--anomalous quantum constraint algebra, therefore one might have to allow for non--zero values. Furthermore in a classical reduction second class constraints lead to the elimination of only half the number of degrees of freedom as for first class constraints. One has to make sure that unphysical gauge degrees do not survive the continuum limit. 

Quantum anomalies might depend on operator orderings and the precise definition of the master constraints but could also arise do to a physical reason, one possibility being a fundamental discretization scale for time like intervals. To differentiate between these issues it might be necessary to quantize the original constraints and to study the commutator algebra on the family of physical Hilbert spaces which arise in taking the continuum limit.

\section{Discussion} \label{laber}

Incorporating diffeomorphism symmetry into quantum gravity models is a very difficult task. It can however ensure the correct semi--classical and continuum limit of quantum gravity. We expect that models involving a discretization of the action will break at least a part of the symmetries to approximate ones, although existent discretizations, as for instance of the Plebanski action, have yet to be checked. One method is to consider the eigenvalues of the Hesse matrix evaluated at a solution. A minimal requirement for discretized actions is that symmetries corresponding to translations of flat vertices are preserved exactly.

This opens the question whether one can construct better actions or discretizations in which all the symmetries are exact. Indeed this is possible for 3d gravity with cosmological constant \cite{lambda}. For 4d gravity renormalization flow methods could yield such (possibly non--local) actions. Lehto et.al.\cite{nielsen} consider an action with quadratic curvature terms in order to force the dynamics to give nearly flat geometries, as for flat geometries the symmetries are exact. In addition symmetries corresponding to a free choice of triangulation could appear in covariant and in canonical formalisms. It would be of interest whether symmetries involving only a change of labels and symmetries involving also a change of triangulation or more general simplicial complex can be related.

On the other hand one can proceed with actions where the symmetries are only approximate. The methods of consistent discretization can be used to obtain canonical theories that reflect the exact and approximate symmetries of such actions. Approximate symmetries lead to approximate constraints -- these are families of constraints labeled by Lagrange multipliers. The algebra of such constraints should be studied and it remains to be seen whether the continuum constraint algebra can be recovered in the limit. At least for discretized hypersurfaces embedded in flat space a first class algebra of constraints can be obtained.

It is necessary to have the same symmetries in the covariant action and the canonical formalism in order to make the idea of a path integral acting as a projector on physical states consistent. To this end also the path integral measure should be taken into account. Here conditions for an approximate invariance have to be formulated. Alternatively a projector on physical states can be constructed using canonical quantization and in most cases some kind of master constraint. Results have to checked carefully for anomalies.

Boundary states \cite{propag} arising from path integrals over a region with boundary should reflect the symmetries of the path integral. The same holds for observables acting on such states. As most of the symmetries are broken this might not constrain states and observables as much as requiring exact gauge invariance. Note however that non--gauge invariant variables might be very sensitive to discretization ambiguities and furthermore reflect unphysical degrees of freedom, as is emphasized in the consistent discretization program \cite{disccosmo}. Indeed in the limit in which the constraints are imposed exactly on the boundary states, operators that do not commute with the constraints will have unbounded fluctuations due to the uncertainty relations. Hence observables commuting either approximately or exactly with the constraints are required. As an exact commutation might restrict the space of observables to much for anomalous constraint algebras, proper conditions have to be formulated. Techniques to compute observables are available for continuum theories \cite{obs} and should be adapted to discrete ones.

\begin{center}{\bf  ACKNOWLEDGMENTS}\end{center}

 It is a pleasure to thank Benjamin Bahr, Laurent Freidel, Renate Loll, Rafael Sorkin, Simone Speziale, Thomas Thiemann, Ruth Williams, the organizer Daniele Oriti and participants of the workshop `Microscopic models of quantum spacetime' for discussions. This work was funded by a Marie Curie Fellowship of the European Union.

\end{document}